\documentclass[12pt]{iopart}

\usepackage{epsfig}
\def\lsim{\raise0.3ex\hbox{$<$\kern-0.75em\raise-1.1ex\hbox{$\sim$}}}
\def\gsim{\raise0.3ex\hbox{$>$\kern-0.75em\raise-1.1ex\hbox{$\sim$}}}
\begin{document}

\title[Charge fluctuations and freeze-out in heavy ion collisions]{Moments of charge fluctuations, pseudo-critical temperatures
and freeze-out in heavy ion collisions}

\author{Frithjof Karsch}

\address{Physics Department, Brookhaven National Laboratory,
Upton, NY 11973, USA;
Fakult\"at f\"ur Physik, Universit\"at Bielefeld, D-33615 Bielefeld, Germany}

\ead{karsch@bnl.gov}

\begin{abstract}
We discuss universal properties of higher order cumulants of net baryon
number fluctuations and point out their relevance for the analysis of 
freeze-out and critical conditions in heavy ion collisions at LHC and RHIC.
\end{abstract}

%\maketitle
\vspace*{-0.1cm}
\section{Chemical freeze-out, the hadron resonance gas and lattice QCD}

Higher order moments, or more accurately higher order cumulants, of conserved
charges are central observables analyzed in the ongoing low energy runs
at RHIC. It also is quite straightforward to analyze their thermal properties
in equilibrium thermodynamics of QCD, e.g. by performing lattice QCD 
calculations. Ratios of cumulants of different order
are particularly well suited for a comparison
with experiment as they are independent of the interaction volume.
Lattice QCD calculations have shown that such ratios, e.g. ratios
of baryon number, electric charge or strangeness fluctuations, are
quite sensitive probes for detecting critical behavior in QCD. They
are sensitive to universal scaling properties at vanishing as well as 
non-vanishing baryon chemical potential ($\mu_B$) and directly reflect
the internal degrees of freedom that are carriers 
of the corresponding conserved charge \cite{cumulantsHRG,cumulantsLGT}
in a thermal medium. 
These ratios change rapidly in the crossover
region corresponding to the chiral transition in QCD and reflect the
change from hadronic to partonic degrees of freedom \cite{lattice}. 
%They thus highlight
%the deconfining features of the QCD transition. 

\begin{figure}[t]
\begin{center}
\epsfig{file=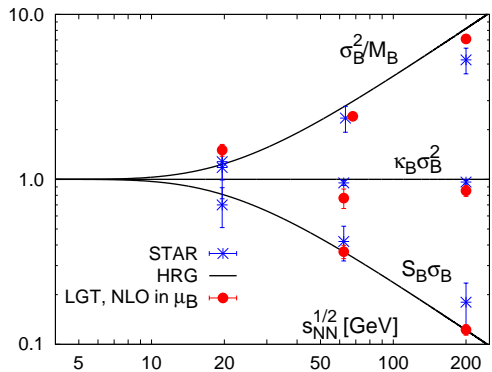,width=74mm}
\end{center}
\vspace{-0.4cm}
\caption{\label{fig:fluctuations} 
Ratios of cumulants of net baryon number fluctuations measured by 
STAR \cite{STAR} and calculated in a HRG model \cite{redlich}. 
The lattice QCD results \cite{Schmidt}
have been determined from a next-to-leading order Taylor expansion of cumulants
calculated at the values of the freeze-out chemical potential determined
from the HRG model. The freeze-out temperature at $\mu_B=0$ has been 
assumed to coincide with the crossover temperature determined in lattice
calculations.
\vspace{-0.3cm}
}
\end{figure}

In the low temperature phase of QCD ratios of cumulants seem to be
well described by a hadron resonance gas model (HRG) 
\cite{cumulantsHRG,lattice,LGTfluctuations}. The HRG is also very 
successful in describing the thermal conditions which characterize the
chemical freeze-out of hadron species. The freeze-out temperature
and its dependence on $\mu_B$ is found to be 
close to the pseudo-critical temperature $T_{pc}$ for the QCD transition. 
We recently pointed out that
also ratios of cumulants of net baryon number fluctuations, as measured 
by STAR at different beam energies \cite{STAR}, agree well with HRG model calculations
on the chemical freeze-out curve \cite{redlich}. This is shown in 
Fig.~\ref{fig:fluctuations}. 
At the same time these results are also consistent with lattice 
QCD calculations when temperature and chemical potential in these
calculations is chosen to agree with the freeze-out parameters
\cite{GG,Schmidt}. 
This also is shown in Fig.~\ref{fig:fluctuations}.

The HRG model, of course, is not sensitive to critical behavior. 
The agreement of current lattice calculations with HRG model calculations
of ratios of cumulants including up to fourth order fluctuations
thus also indicates that they present so far no compelling evidence for 
critical behavior. 
%A natural question to ask thus is to what extent
%fluctuations of conserved charges can help to characterize the freeze-out
%conditions in heavy ion collisions and at the same time can help
%to extract information on the QCD phase diagram. 
We recently argued that a determination of sixth order cumulants 
may be particularly helpful in this 
respect as they clearly deviate from HRG model calculations in the
crossover region of QCD and start to become sensitive to critical 
behavior earlier in the hadronic phase \cite{Friman}. In the following we will
point out some generic features  of higher order cumulants in QCD that
arise from universal properties of QCD close to the chiral limit of
vanishing light quark masses, $m_q\equiv m_u=m_d =0$.

\section{O(4) universality}

\begin{figure}[t]
\begin{center}
\hspace*{0.4cm}\epsfig{file=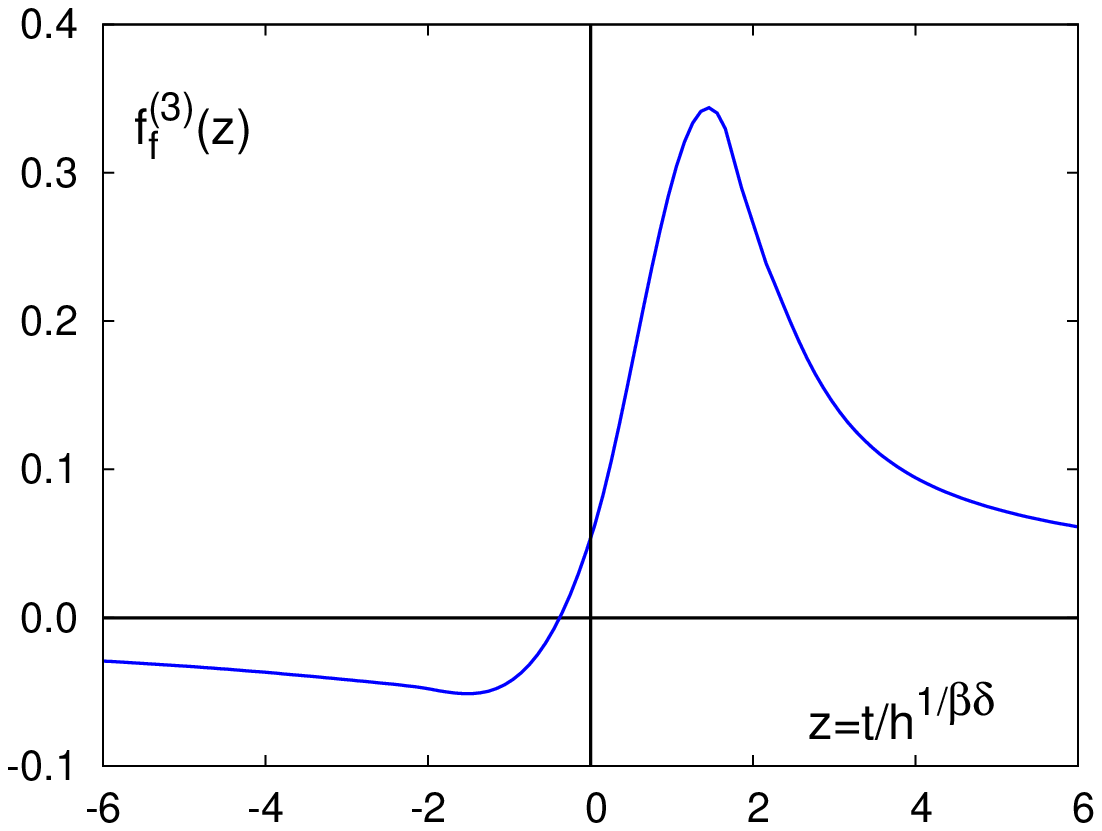,height=55mm}\hspace*{0.4cm}
\epsfig{file=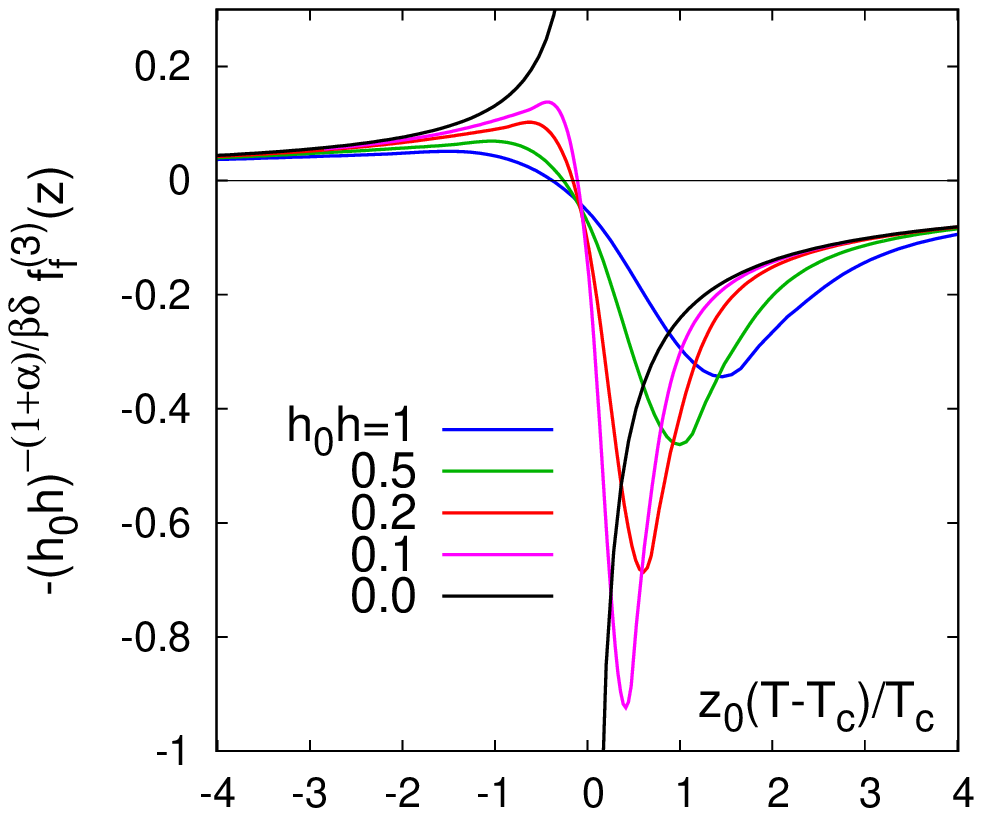,height=55mm}
\end{center}
\vspace{-0.4cm}
\caption{\label{fig:scaling}
The third derivative of the singular part of the free energy in theories
belonging to the 3-d, O(4) universality class (left) and its 
contribution to third or higher order cumulants (see Eq.~2)
(right). $h_0$ and $z_0$ are non-universal scale parameters.
\vspace{-0.3cm}
}
\end{figure}

\vspace{-0.2cm}
At vanishing baryon chemical potential ($\mu_B$), as well as in a 
certain range $0\le \mu_B\le \mu_B^c$, QCD is expected to 
undergo a second order phase transition for  
$m_q\equiv 0$. In this entire range of $\mu_B$ values the 
chiral phase transition belongs to the universality class of 
3-dimensional, O(4) symmetric spin
models. In the vicinity of the chiral phase transition temperature
($T_c (\bar{\mu}_B)$) 
thermodynamic quantities show universal properties that are controlled
by the singular part, $f_f$, of the free energy. For a fixed value
of the chemical potential $\bar{\mu}_B<\mu_B^c$, close to 
$T_c (\bar{\mu}_B)$ the free energy may be parametrized as, 
\begin{equation}
f(T,\mu_B,m_q) = h^{2-\alpha} f_f(z) + f_r(T,\mu_B,m_q) \; ,
\label{freeenergy}
\end{equation}
with $z\equiv t/h^{1/\beta\delta}$ denoting the particular
scaling combination of the reduced temperature
$t\sim (T- T_c(\bar{\mu}_B))/T_c(\bar{\mu}_B) + \kappa_B \left( 
(\mu_B/T)^2- (\bar{\mu}_B/T)^2\right)$
and the symmetry breaking parameter $h$, which for QCD is taken to be
the ratio of light and strange quark masses, $h\sim m_q/m_s$;
$\alpha,\ \beta, \ \delta$ 
denote critical exponents of the 3-d, O(4) universality class. 
Cumulants of net baryon number
fluctuations are then obtained as derivatives of $f(T,\mu_B,m_q)$ with
respect to $\hat{\mu}_B \equiv \mu_B/T$, i.e., 
$\chi_n^{B} \sim \partial^n f/\partial \hat{\mu}_B^n$.
Close to the chiral limit and for $n\ge 3$ these derivatives are dominated 
by the singular part,
\begin{equation}
\chi_n^B \sim
%\begin{cases}
\cases{
- m_q^{(2-\alpha -n/2)/\beta\delta} f_f^{(n/2)}(z)
& ,\ {\rm for}\ $\bar{\mu}_B /T = 0$,\
{\rm and}\ $n$ \ {\rm even} \\
- \left( \frac{\bar{\mu}_B}{T} \right)^n
m_q^{(2-\alpha -n)/\beta\delta} f_f^{(n)}(z)
&,\ {\rm for}\ $\bar{\mu}_B/T > 0$ \,
}
%\end{cases}
\label{fluct_mass}
\end{equation}
The first derivative of the scaling function that leads in the chiral limit
to a divergent cumulant at $T_c(\bar{\mu}_B)$ and, in fact, leads to a 
change of sign of $\chi_n^B$ at $T_c(\bar{\mu}_B)$ is obtained from
the third derivative of the scaling function $f_f^{(3)}$. In 
Fig.~\ref{fig:scaling} (left) we show this scaling function for the 
3-d, O(4) universality class which recently has been extracted from high
statistics spin model calculations \cite{Engels:2011km}. 
The right hand part of this figure
shows how the singular contribution strengthens as the symmetry breaking parameter
$h$ is reduced. 

It is evident from Fig.~\ref{fig:scaling}(right) that the universal 
scaling properties of cumulants will, for sufficiently small values of the 
quark mass, induce negative values for cumulants already for $n=3$. 
However, as shown in Eq.~\ref{fluct_mass}, the contribution from the
singular part is weighted by a factor proportional to $(\mu_B/T)^3$. Its
contribution to the total value of the cumulant may thus be small for 
small values of $\mu_B/T$ and non-zero values of the quark mass. Nonetheless,
negative values for $3^{rd}$ and $4^{th}$ order moments have been found in
model calculations \cite{Friman,Asakawa}\footnote{Close to the chiral critical point
the negative values of fourth order cumulants that arise from O(4) 
criticality compete with similar effects that arise from Z(2) critical
behavior and also lead to negative fourth order cumulants 
\cite{Stephanov}.}

\begin{figure}[t]
\begin{center}
\epsfig{file=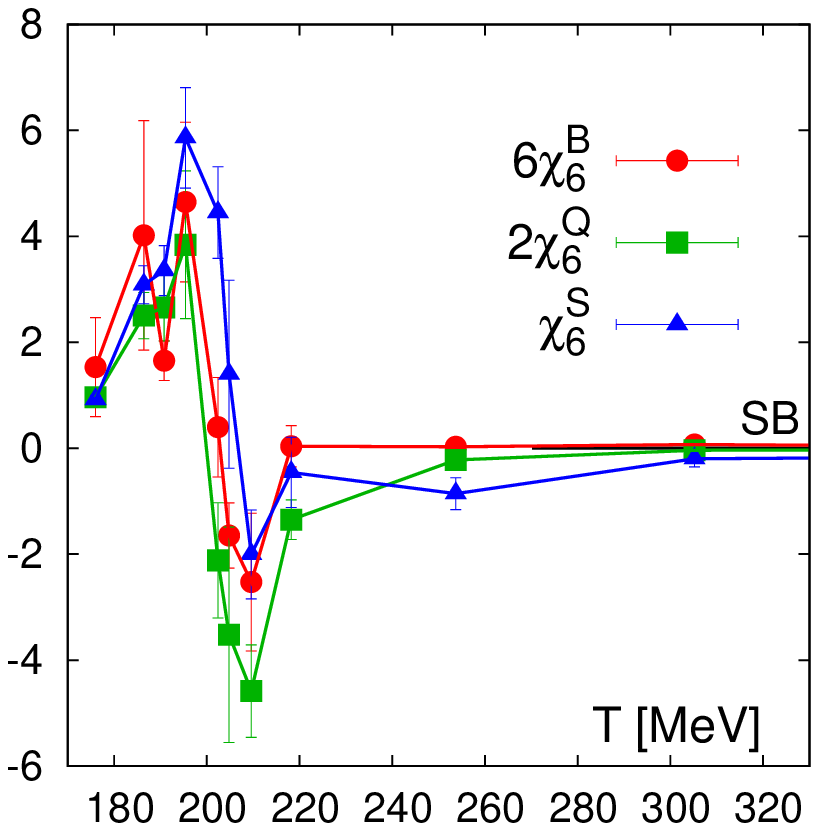,height=54mm}
\epsfig{file=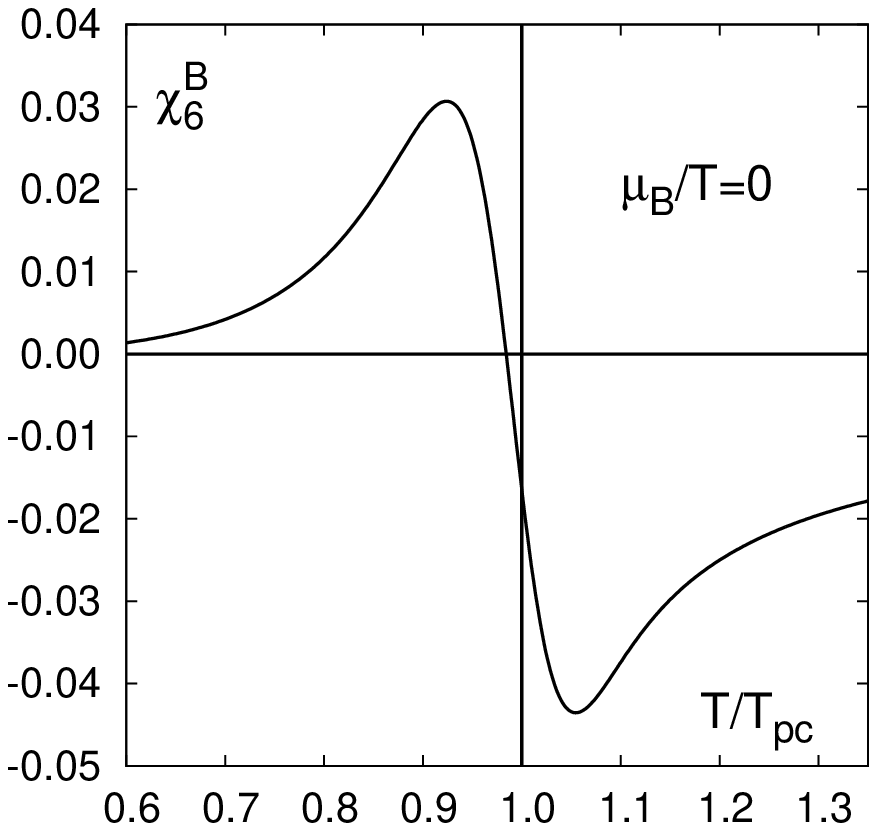,height=54mm}
\end{center}
\vspace{-0.4cm}
\caption{\label{fig:c6}
Sixth order cumulants of net baryon number, electric charge
and strangeness fluctuations calculated in (2+1)-flavor QCD at $\mu_B=0$
\cite{LGTfluctuations,Schmidt} (left) and in the PQM model (right) 
\cite{Friman}.
\vspace{-0.3cm}
}
\end{figure}

The first cumulant, which is not parametrically suppressed by powers of 
$\mu_B/T$ and thus stays finite even at $\mu_B=0$ is the sixth order 
cumulant of net baryon number fluctuations. This cumulant changes sign in 
the crossover
region of the QCD transition. Although the temperature at which
the sixth order cumulant changes sign is not a universal quantity there are
strong indications from lattice \cite{LGTfluctuations} as well as model 
\cite{Friman} calculations that the change of sign occurs close to the 
chiral transition temperature.
This is shown in Fig.~\ref{fig:c6}. 
In fact, for sufficiently small 
values of the quark mass, i.e., in the O(4) scaling regime, the location of
the minimum of the sixth order cumulant and its position relative to the
pseudo-critical temperature for the chiral transition is controlled by the 
location of the maximum of $f_f^{(3)}$ and its location relative to the 
peak in the chiral susceptibility. The latter appears at a somewhat higher
temperature \cite{Engels:2011km}. For $\mu_B/T > 0$ the onset of negative
values for $\chi_6^B$ indeed follows the crossover line for the QCD
transition \cite{Friman,Skokov}.

\vspace*{-0.2cm}
\section{Conclusions}
\vspace*{-0.1cm}
Sixth order cumulants are thus expected to 
change sign at a temperature below the (pseudo-critical) chiral transition 
temperature.  This effect should become visible even at the LHC and the
highest RHIC energy if chemical freeze out indeed occurs close to the QCD 
transition temperature and if higher moments probe these freeze out conditions.

\vspace*{-0.2cm}
\section*{Acknowledgments}
\vspace*{-0.2cm}
This manuscript has been authored under contract number
DE-AC02-98CH10886 with the U.S. Department of Energy.


\vspace*{-0.2cm}
\section*{References}
\vspace*{-0.2cm}
\begin{thebibliography}{10}

\bibitem{cumulantsHRG}
F.~Karsch, K.~Redlich, A.~Tawfik,
  %``Thermodynamics at nonzero baryon number density: A Comparison of lattice and hadron resonance gas model calculations,''
  Phys.\ Lett.\  {\bf B571}, 67-74 (2003),
  [hep-ph/0306208].
\bibitem{cumulantsLGT}
  C.~R.~Allton et al., 
%M.~Doring, S.~Ejiri, S.~J.~Hands, O.~Kaczmarek, F.~Karsch, E.~Laermann, K.~Redlich,
  %``Thermodynamics of two flavor QCD to sixth order in quark chemical potential,''
Phys.\ Rev.\  {\bf D71}, 054508 (2005),
[hep-lat/0501030].
\bibitem{lattice}
S.~Ejiri, F.~Karsch and K.~Redlich,
%``Hadronic fluctuations at the QCD phase transition,''
Phys.\ Lett.\  {\bf B633}, 275-282 (2006).
\bibitem{LGTfluctuations}
M.~Cheng et al., 
%P.~Hegde, C.~Jung, F.~Karsch, O.~Kaczmarek, E.~Laermann, R.~D.~Mawhinney, C.~Miao {\it et al.},
  %``Baryon Number, Strangeness and Electric Charge Fluctuations in QCD at High Temperature,''
  Phys.\ Rev.\  {\bf D79}, 074505 (2009),
  [arXiv:0811.1006 [hep-lat]].
\bibitem{STAR}
M.~M.~Aggarwal et al. (STAR Collaboration), Phys. Rev. Lett. {\bf 105}
(2010) 22302.
%arXiv:1004.4959v2 [nucl-ex].
\bibitem{redlich}
F.~Karsch, K.~Redlich,
  %``Probing freeze-out conditions in heavy ion collisions with moments of charge fluctuations,''
Phys.\ Lett.\  {\bf B695}, 136 (2011),
[arXiv:1007.2581 [hep-ph]].
\bibitem{GG}
R.~V.~Gavai, S.~Gupta,
  %``Lattice QCD predictions for shapes of event distributions along the freezeout curve in heavy-ion collisions,''
  Phys.\ Lett.\  {\bf B696}, 459-463 (2011).
\bibitem{Schmidt}
C.~Schmidt,
  %``Net-baryon number fluctuations in (2+1)-flavor QCD,''
  Prog.\ Theor.\ Phys.\ Suppl.\  {\bf 186}, 563-566 (2010),
  [arXiv:1007.5164 [hep-lat]].
\bibitem{Friman}
  B.~Friman, F.~Karsch, K.~Redlich, V.~Skokov,
  %``Fluctuations as probe of the QCD phase transition and freeze-out in heavy ion collisions at LHC and RHIC,''
Eur.\ Phys.\ J.\  {\bf C71}, 1694 (2011).
%[arXiv:1103.3511 [hep-ph]].
\bibitem{Engels:2011km}
  J.~Engels, F.~Karsch,
  %``The scaling functions of the free energy density and its derivatives for the 3d O(4) model,''
  [arXiv:1105.0584 [hep-lat]].
\bibitem{Asakawa}
M.~Asakawa, S.~Ejiri and M.~Kitazawa,
%``Third moments of conserved charges as probes of QCD phase structure,''
Phys.\ Rev.\ Lett.\  {\bf 103}, 262301 (2009).
\bibitem{Stephanov}
M.~A.~Stephanov,
  %``On the sign of kurtosis near the QCD critical point,''
  [arXiv:1104.1627 [hep-ph]].
\bibitem{Skokov}
V. Skokov, these proceedings.
\end{thebibliography}
\end{document}